\documentstyle[epsf]{l-aa}
\begin{document}
\thesaurus{09(19.82.1)}
\title{Influence of heavy element and rotationally induced diffusions on the 
solar models}
\author{M. Gabriel}
\offprints{M. Gabriel}
\institute{Institut d'Astrophysique de l'Universit\'e de Li\`ege, 
5, avenue de Cointe, B-4000 Li\`ege, Belgium.}
\date{Received  ,Accepted  }
\maketitle 
\begin{abstract}
We present the results of solar model computation done with the latest
Livermore opacities. Models without diffusion, with hydrogen diffusion only and
with hydrogen and heavy elements diffusion are considered. The influence of
mixing below the convective envelope induced by rotation and angular momentum
losses is also discussed. The sound speed of our best model,  does 
not deviate from that of Basu's seismic one by more than about $10^{-3}$;  
p-mode frequencies are also compared with observations and frequencies for low
 order p and g-modes are given
\end{abstract}

\section{Introduction}
Solar models have improved significantly over the past years mainly due to
improvements in the physics. Recently Iglesias \& Rogers (1996) have published 
new opacities which we have used to compute new improved solar models.
 In addition, 
 we have included the diffusion of the heavy elements in our code. We
think interesting to present a set of models with the 3 possible hypotheses
concerning diffusion (no, hydrogen only, H and Z diffusion) and having for the
 rest exactly the same input physics. 
These models show too much gravitational settling below the convective envelope.
To correct this situation a mixing induced by rotation and angular momentum
losses is introduced. This leads to a significant improvement in the fit of the
models with the seismic one but now the theoretical and observed helium values
 slightly disagree but the difference remains within the observational error.

In Sect. 2 we give the input physics and the sources for our comparisons. In 
Sect. 3 we present and discuss the results.

\section{Input physics}
\begin{table}
\caption{Most abundant ions and their fraction $x$ at $T=2.22 10^6$ and at 
the center of the Sun}
\begin{tabular}{|c|c|c|c|c|} \hline
\multicolumn{1}{|c|}{} & \multicolumn{2}{c|}{ $T=2.22 10^6$} &
\multicolumn{2}{c|}{center} \\ \hline
element & ion & $x$ & ion & $x$  \\ \hline
$C^{12}$ & $C_{VII}$ & 0.885 & $C_{VII}$ & 1. \\ 
$N^{14}$ & $N_{VIII}$ & 0.753 & $N_{VIII}$ & 1. \\
$O^{16}$ & $O_{IX}$ & 0.502 & $O_{IX}$ & 0.997 \\
$Ne^{20}$ & $Ne_{IX}$ & 0.514 & $Ne_{XI}$ & 0.926 \\
$Mg^{24}$ & $Mg_{XI}$ & 0.912 & $Mg_{XIII}$  & 0.727 \\
$Si^{28}$ & $Si_{XIII}$ & 0.894 & $Si_{XV}$ & 0.485\\
$S^{32}$ & $S_{XV}$ & 0.760 & $S_{XVI}$  & 0.470 \\
$Fe^{56}$ & $Fe_{XVIII}$ & 0.542 & $Fe_{XXV}$ & 0.851 \\
\hline 
\end{tabular}
\end{table}
\begin{table*}
\caption{Main properties of the models. $X_0$ ($Z_0$) and $X_S$ ($Z_S$) are  
respectively the initial and the final hydrogen (heavy element) surface 
abundances . The subscripts c and e refer 
to the center and to the bottom of the convective envelope respectively. $q_e$ 
and $x_e$ are respectively the mass fraction and the fractional radius  at the 
bottom of the envelope.  Model 1: no diffusion, model 2: hydrogen diffusion only,
 models 3 and 4: hydrogen and heavy element diffusions, models 5 to 7: also 
rotationally induced mixing}
\begin{flushleft}
\begin{tabular}{|c|c|c|c|c|c|c|c|c|c|c|c|c|} \hline
Nr & $X_0$ & $X_S$ & $X_c$ & $Z_0$ & $Z_S $ & $Z_c$ &$\rho_c$ & $T_c$ 
$10^{-6}$  & $\rho_e$ & $T_e$$10^{-6}$ & $q_e$ & $ x_e$ \\ \hline 
1 & 0.701737 & 0.70174 & 0.34658 & 0.02 & 0.02 & 0.02 & 149.48 & 15.584 & 
0.16791 & 2.1411 & 0.978533 & 0.72321 \\ 
2 & 0.704093 & 0.73566 & 0.32819 & 0.02 & 0.02 & 0.02 & 154.64 & 15.682 &
0.19869 & 2.2523 & 0.974144 & 0.70760 \\
3 & 0.704644 & 0.73688 & 0.33051 & 0.02 & 0.01801 & 0.02091 & 153.87 & 15.676 
& 0.18726 & 2.1861 & 0.975775 & 0.71360 \\
4 & 0.699316 & 0.73174 & 0.32392 & 0.021 & 0.01893 & 0.02179 & 154.49 & 15.744 
& 0.19270 & 2.2158 & 0.975021 & 0.71159 \\
5 & 0.704617 & 0.72840 & 0.33039 & 0.020 & 0.01864 & 0.02081 & 153.91 & 15.674 
& 0.18719 & 2.1931 & 0.9758034 & 0.71424 \\
6 & 0.704623 & 0.72698 & 0.33041 & 0.020 & 0.01875 & 0.02080 & 153.90 & 15.674 
& 0.18723 & 2.1946 & 0.975800 & 0.71431 \\
7 & 0.701942 & 0.72663 & 0.32714 & 0.0205 & 0.01908 & 0.02127 & 154.20 & 15.709 
& 0.19003 & 2.2082 & 0.975407 & 0.71308 \\
\hline 
\end{tabular}
\end{flushleft}
\end{table*}

May be, it is worth recalling that our equation of state takes into account the 
ionization of the 8 most abundant heavy elements (those considered in table 1) but 
does not include their excited states in the computation of the internal partition 
function. It includes the Deb\"{y}e correction modified as in Gabriel (1994b) 
with his parameter $\alpha=0.1$.  
The input physics is the same as in our previous papers (Gabriel 1994a,b, 1995,
1996; Gabriel \& Carlier 1997) except for 2 points. We now use the latest 
Livermore opacities (Iglesias \& Rogers 1996) and their interpolation routine 
still complemented at low temperatures by those of Neuforge (1993). We have 
included  diffusion using Thoul et al. (1994) theory. In Gabriel \& Carlier 
(1997), we used Thoul et al.'s interpolation formulae to compute the coefficients 
$Ap$, 
$At$ and $Ax$, now we call Thoul et al.'s routine to compute these coefficients. 
The diffusion theories used in stellar evolution computations suffer from 
several
uncertainties. One of them comes from the use of the Debye shielding for the
computation of the cross-section for Coulomb scattering as, in the Sun, the 
Debye radius is only a slightly larger than the mean distance between particles. 
However the main uncertainty comes from the hypothesis of full ionization 
which is not valid for the heavy elements. Table 1 gives the ionization 
state of the most abundant heavies close to the bottom of the convective zone 
(at $T= 2.22 10^6$ K) and at the center of the Sun. It shows that none of them 
is fully ionized everywhere below the convective envelope and that most of 
them are nowhere fully ionized in the Sun. In such circumstances the results 
obtained with the diffusion of the heavy elements must be considered with 
caution and it is useless to make  detailed computations. Also we have assumed 
that all heavies diffuse at the same rate given by that of $O^{16}$. 
Nevertheless the  abundances of 
the CNO elements are computed taking both nuclear reactions and diffusion 
simultaneously into account using the same method as in Gabriel \& Carlier 
(1997) for hydrogen burning.

These models with diffusion show too much gravitational settling below the 
convective envelope. This can have two causes. 
Since most of the heavy elements are not fully ionized there, the theory could
predict wrong diffusion rates leading to an underestimate of the opacity 
 but there is nothing we can do to correct this point as a new,
much complex theory, would be required. However, since the model with hydrogen 
diffusion only also shows this behaviour, it is likely that another process which 
has been neglected is at work. It is known that mixing must occur in that region
because of the angular momentum diffusion and of the shape of the rotation
law which is very latitude dependent (see for instance Corbard et al. 1997).
 Mixing is also required to explain the $Li^7$ depletion (see of instance 
Richard et al. 1996). 
There are many estimates of the diffusion coefficient connected to 
rotation (see Pinsonneault 1996 for a review of this problem) but the most 
reliable is provided by the numerical results obtained with the
Yale code (Pinsonneault et al. 1989, 1990). An expression for the diffusion
coefficient fitting some of their results has been given by Proffitt \& 
Michaud (1991).
We have used their formula with slight modifications to take that effect into
account. 

As usual, all the evolutions are started during the gravitational contraction as
early as allowed by the extend of the Livermore opacity tables. An iteration 
on the initial hydrogen abundance and on the mixing length is used to get 
the solar radius and luminosity to better than $10^{-5}$. The solar age is
assumed to be equal to $4.6 10^9$ years. 

The observed value of $Z/X= 0.0245$  has an uncertainty which might be of the 
order of 10\%  because the most abundant heavy elements (the CNO group and Ne) 
show very few nice, not blended, line and that their f-values are only known 
theoretically  (Grevesse 1997). With $X \simeq 0.73$, the surface $Z$ value can 
be anywhere between 0.016 and 0.02 which does not constrain significantly the 
models. For models without diffusion and with hydrogen diffusion only, we have
taken $Z=0.02$ because this value has been known for long to give a better fit
with observations than that which would be deduced from the observed $Z/X$.
This result may be considered as an indication that $Z$ in most of the
radiative zone is close to 0.02 (Basu et al. 1996, Gabriel \& Carlier 1997). 
For models 
with heavy elements diffusion several initial values of $Z$ have been used. 

Helioseismology provides much stronger constrains on the models. First a large
number of frequencies are known in the 5 min. range but also inversions have
led to seismic models, to informations on the depth of the convective zone
and on the surface helium abundance. Here we use Basu's inverse model (Basu et
al. 1996) and the frequencies given by Chaplin et al. (1996) and Elsworth et al. 
(1994) complemented by those of Anguera Gubau et al. (1992) for degrees smaller 
than 4, for higher ones we use Libbrecht et al. (1990) data. The bottom of the 
convective envelope is located at $x= 0.713 \pm 
0.001$ according to  Basu \& Antia (1997) while Dziembowski et al. (1994) give 
$x=0.71455 \pm 0.00025$. Many values have been given for the helium abundance 
(see Berthomieu 1996 for a discussion of this point). The latest one has been
obtained by Basu \& Antia (1995) who give $Y=0.2456 \pm 0.0007$ for the MHD 
 equation of state and $Y=0.2489 \pm 0.0028$ for OPAL.

\begin{figure*}
 \picplace{7.cm}
\includegraphics{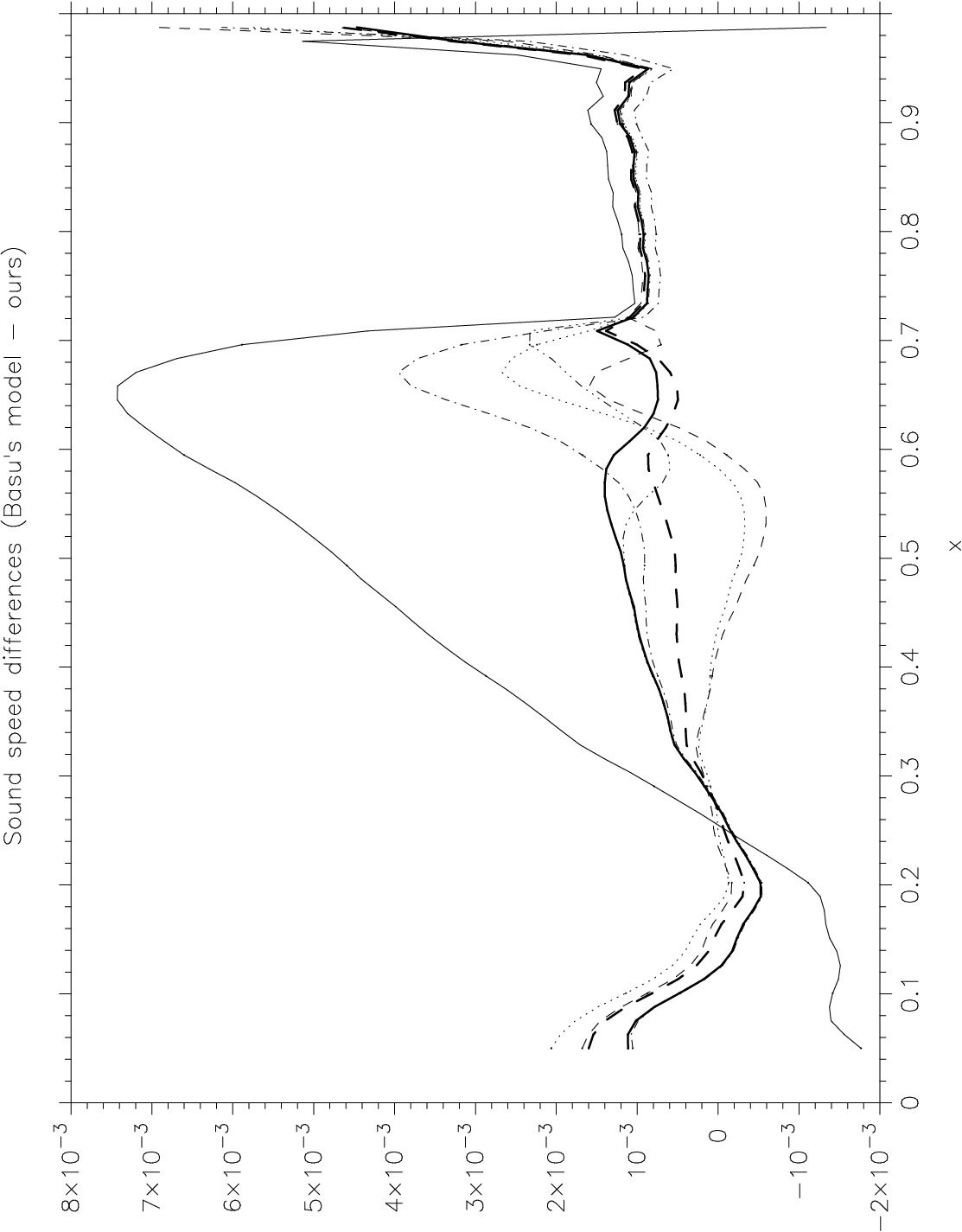}
\caption{Relative difference between the sound speed of Basu's seismic model
and ours. Models 1 to 5 correspond respectively to full, dashed, dot-dash, 
dotted, dash-dot-dot-dot lines, models 6 and 7 to full and dashed heavier 
lines } 
\end{figure*}

\begin{figure*}
 \picplace{7.cm}
\includegraphics{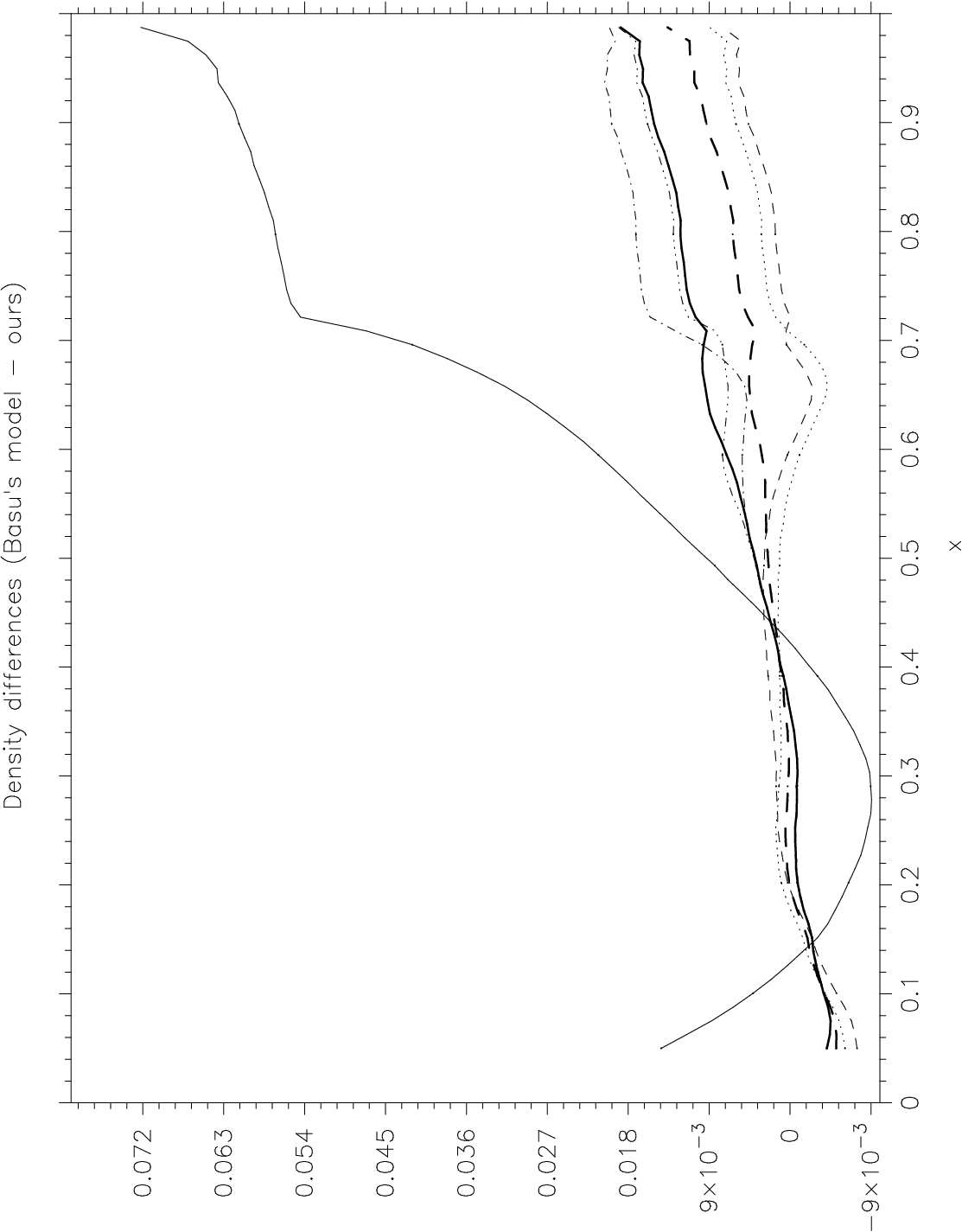}
\caption{Relative difference between the density of Basu's seismic model 
and ours. For conventions see Fig. 1 } 
\end{figure*}
\begin{figure*}
 \picplace{7.cm}
\includegraphics{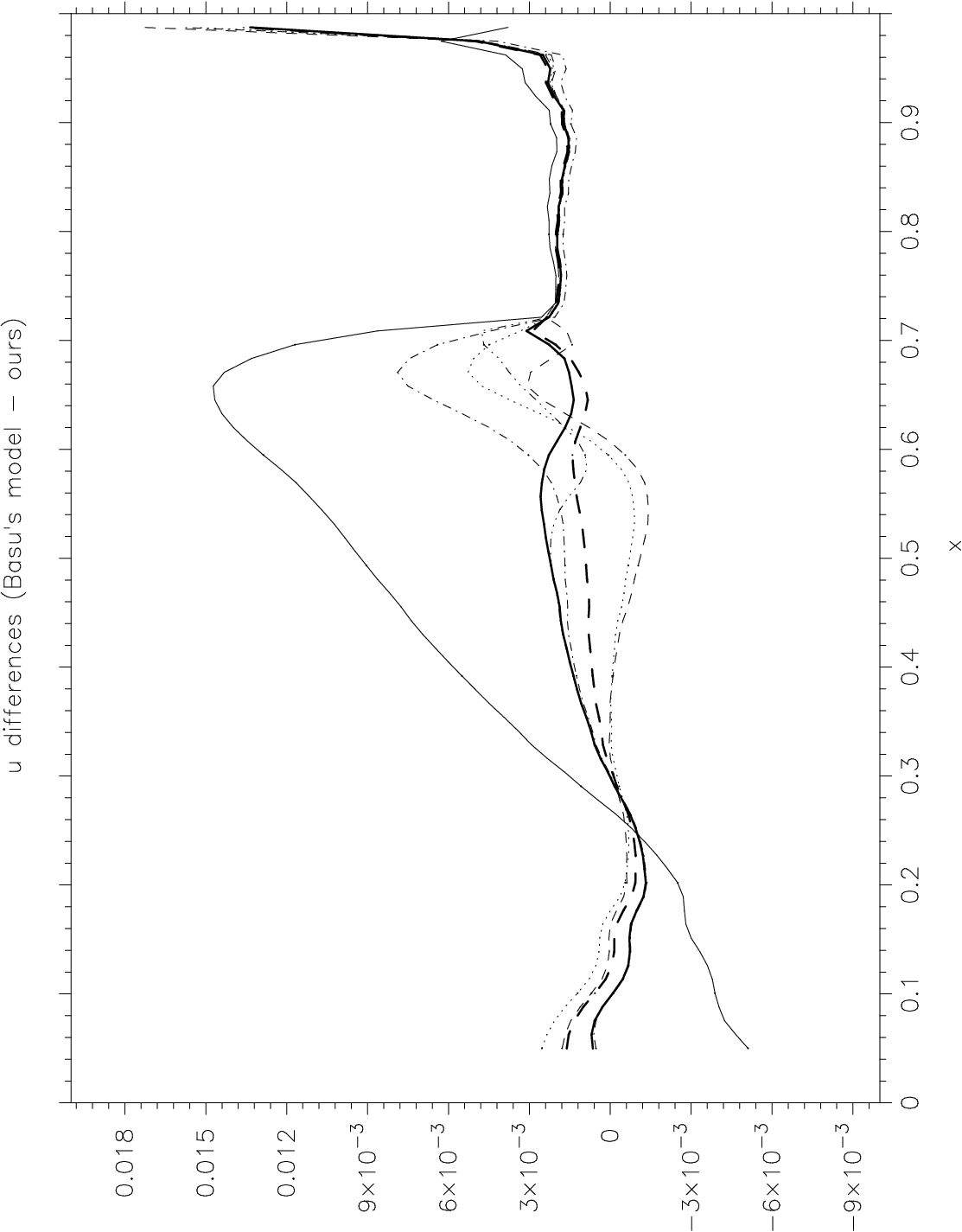}
\caption{Relative difference between $u=p/\rho$ of Basu's seismic model 
and ours. For conventions see Fig. 1} 
\end{figure*}

\section{The results}
\begin{figure}
 \picplace{7.cm}
\includegraphics{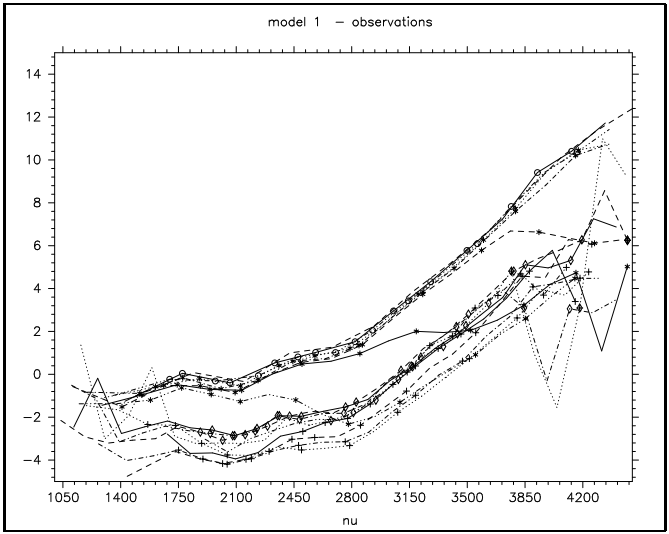}
\caption{Normalized frequencies of model 1 (no diffusion) minus observed ones 
for degrees between 0 and 100.} 
\end{figure}
\begin{figure}
 \picplace{7.cm}
\includegraphics{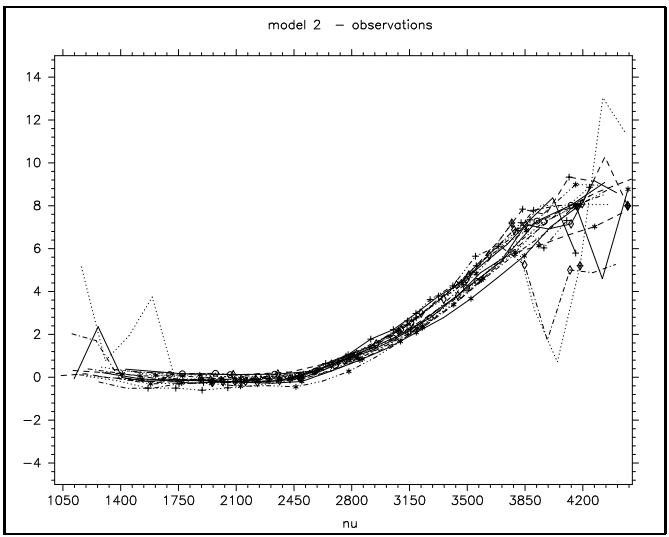}
\caption{Normalized frequencies of model 2 (hydrogen diffusion only) minus 
observed ones for degrees between 0 and 100.} 
\end{figure}
\begin{figure}
 \picplace{7.cm}
\includegraphics{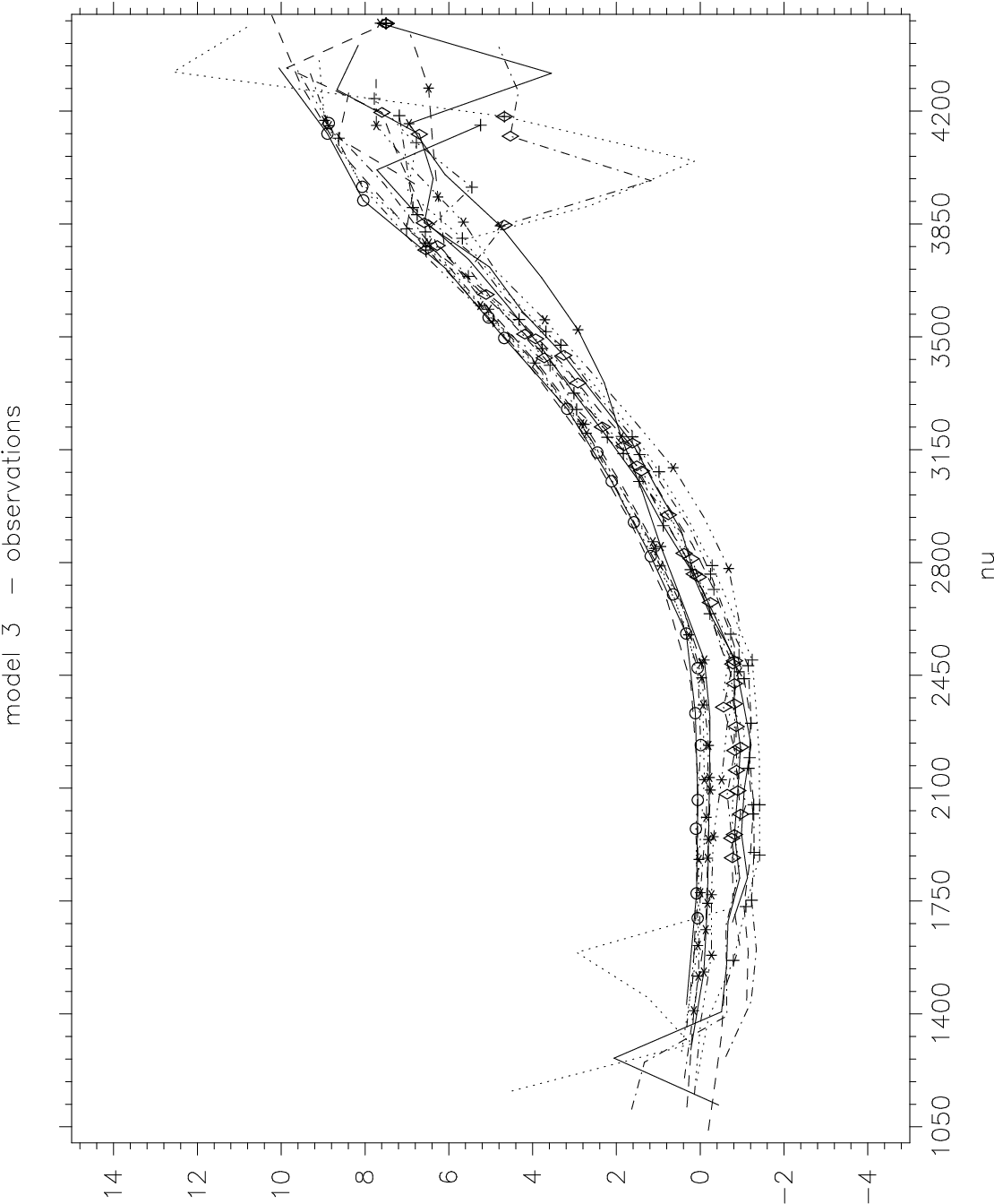}
\caption{Normalized frequencies of model 3 (diffusion of X and Z, $Z_0=0.02$) 
 minus observed ones for degrees between 0 and 100.} 
\end{figure}
\begin{figure}
 \picplace{7.cm}
\includegraphics{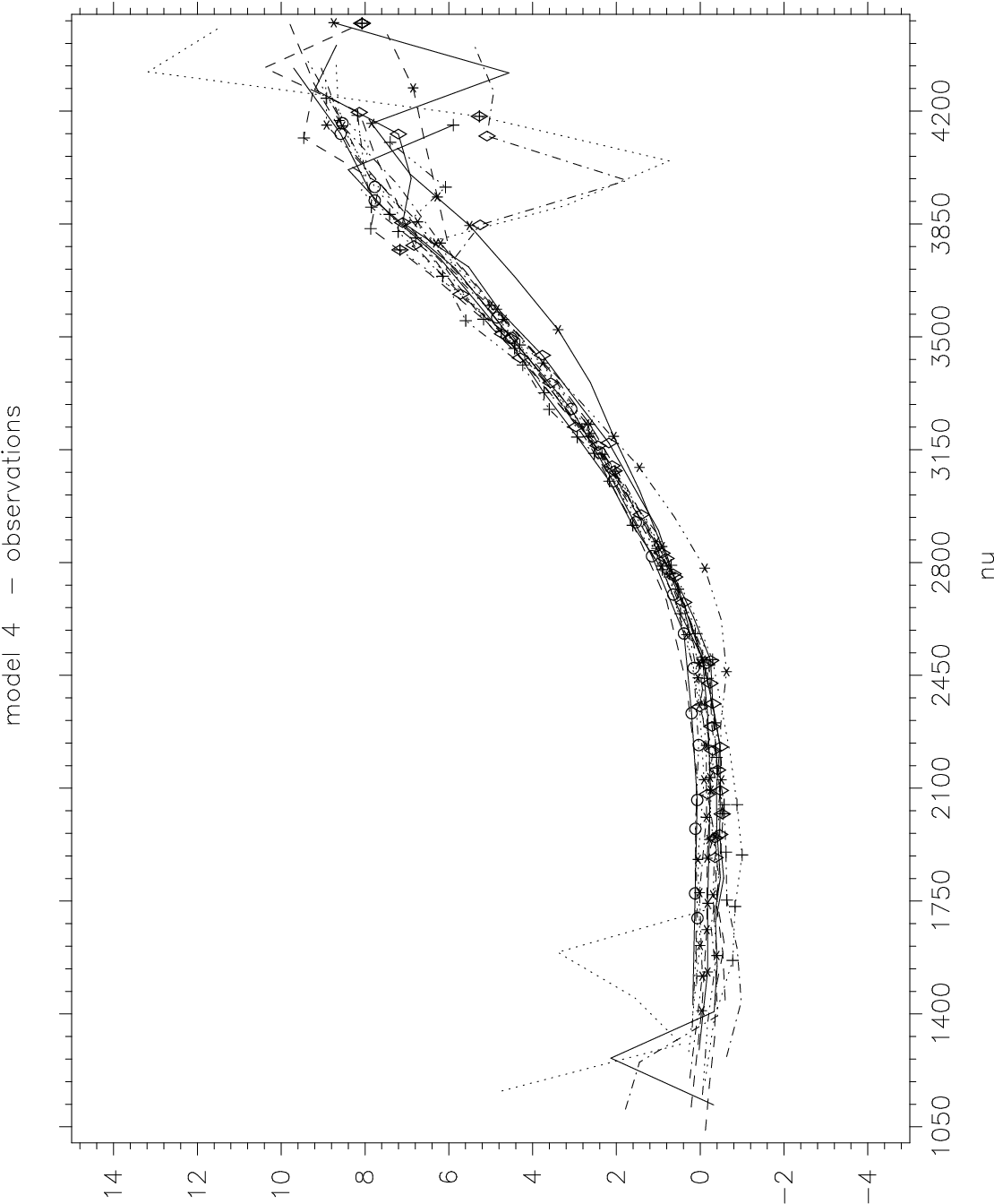}
\caption{Normalized frequencies of model 4 (diffusion of X and Z, $Z_0=0.021$) 
 minus observed ones for degrees between 0 and 100.} 
\end{figure}
\begin{figure}
 \picplace{7.cm}
\includegraphics{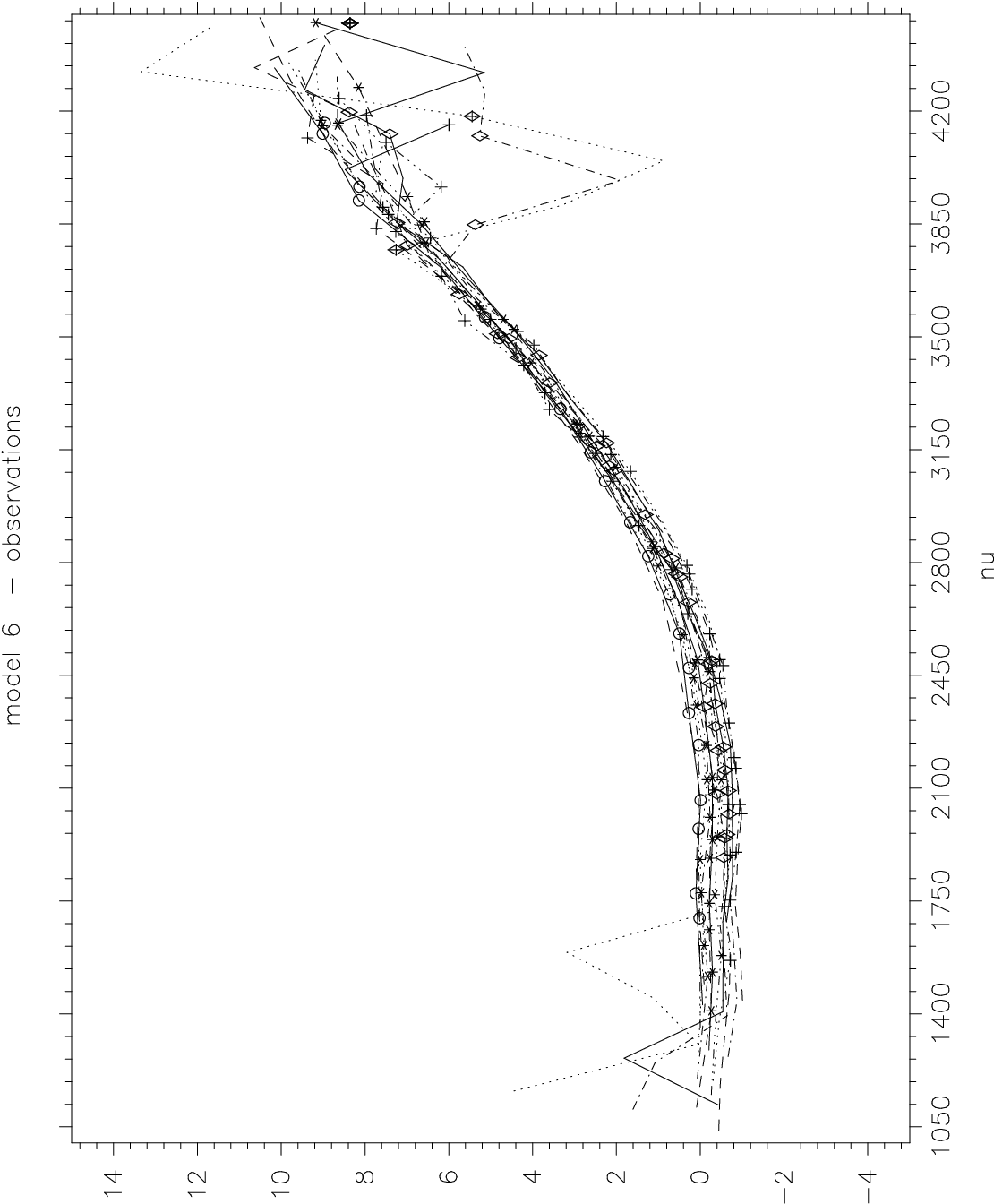}
\caption{Normalized frequencies of model 6 (model with rotationally induced
mixing and $Z_0=0.02$) minus observed ones for degrees between 0 and 100.} 
\end{figure}
\begin{figure}
 \picplace{7.cm}
\includegraphics{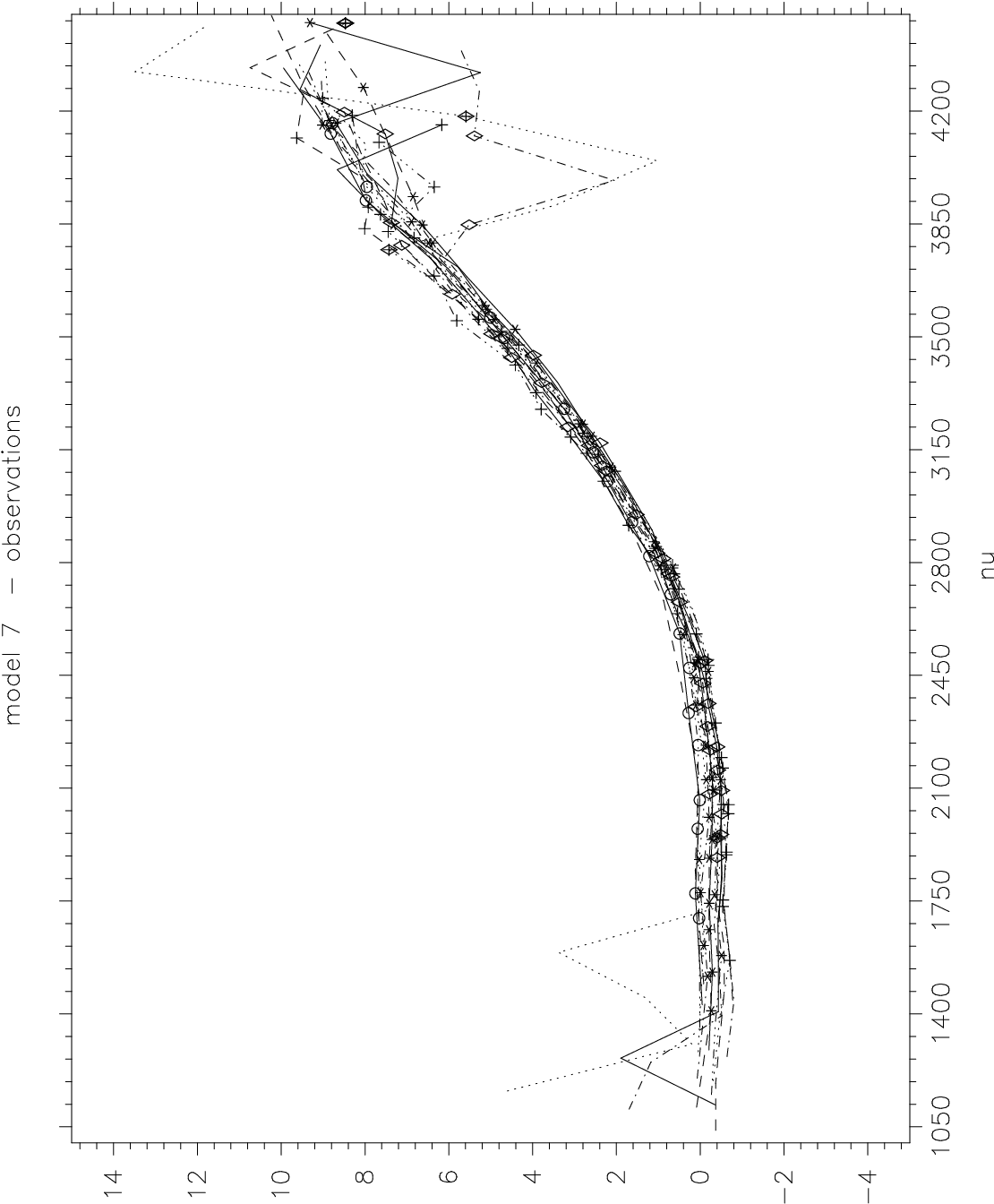}
\caption{Normalized frequencies of model 7 (model with rotationally induced
mixing and $Z_0=0.0205$) minus observed ones for degrees between 0 and 100.} 
\end{figure}
\begin{figure}
 \picplace{7.cm}
\includegraphics{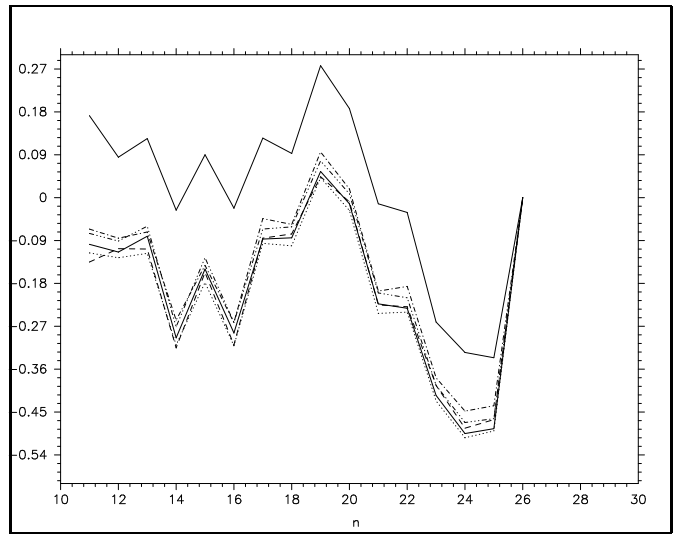}
\caption{Difference between the theoretical and observational values of $d_0(n)$.
 Models 1 to 4 and 6 correspond respectively to full, dashed, dot-dash, 
dotted, dash-dot-dot-dot lines, model 7 to full heavier lines }
\end{figure}
\begin{figure}
 \picplace{7.cm}
\includegraphics{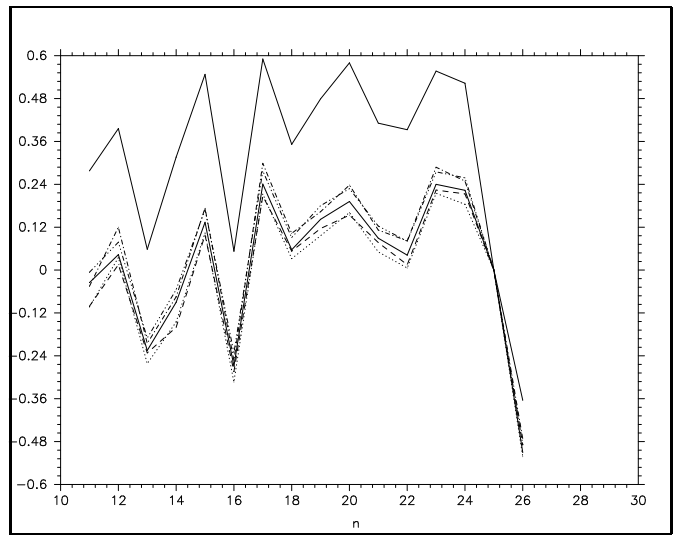}
\caption{Difference between the theoretical and observational values of $d_1(n)$. 
Conventions are the same as for Fig. 10}
\end{figure}
The main properties of the models are summarized in table 2. Model 1 does not
take diffusion into account, model 2 has hydrogen diffusion only, while 
diffusion of heavy elements is included in models 3 and 4 with initial $Z$ 
values of 0.02 and 0.021 respectively. Model 5 includes the diffusion of
hydrogen and heavy elements and the rotationally induced mixing according to
Proffitt \& Michaud (1991) formula. For model 6, their formula was slightly 
modified.
  Model 7 gives our best solar model.
 
The comparison of these models with Basu's seismic model is 
given in Fig. 1, 2 and 3 respectively for the sound speed, the density and
$u=p/\rho$.  
Fig. 4 to 9 give the normalized differences between theoretical and 
observed frequencies.
Fig. 10 and 11 compare the theoretical and observed fine-structure 
spacings.  
\begin{table*}
\caption{Low order p-mode and g-mode frequencies (in Hz) of model 7 
for degrees 0 to 4, n is the order and is negative for g-modes. }
\begin{center}
\begin{tabular}{|c|c|c|c|c|c|} \hline
n & $l=0$ & $l=1$ & $l=2$ & $l=3$ & $l=4$  \\ \hline 
 10 & 1.547916 $10^{-3}$ & 1.612160 $10^{-3}$& 1.674057 $10^{-3}$& 1.728650 $10^{-3}$& 1.777524 $10^{-3}$  \\ 
  9 & 1.407034 $10^{-3}$ & 1.472315 $10^{-3}$ & 1.472315 $10^{-3}$& 1.591030 $10^{-3}$& 1.640368  $10^{-3}$ \\
  8 & 1.262960 $10^{-3}$ & 1.329214 $10^{-3}$& 1.394182 $10^{-3}$& 1.450456 $10^{-3}$ & 1.499757 $10^{-3}$ \\
  7 & 1.117858 $10^{-3}$ & 1.185208 $10^{-3}$& 1.250203 $10^{-3}$& 1.306318 $10^{-3}$ &  1.355870 $10^{-3}$\\
  6 & 9.725076 $10^{-4}$ & 1.039211 $10^{-3}$ & 1.104895 $10^{-3}$& 1.161399 $10^{-3}$ &  1.210142 $10^{-3}$ \\
  5 & 8.252254 $10^{-4}$ & 8.936510 $10^{-4}$ & 9.597224 $10^{-4}$& 1.014746  $10^{-3}$& 1.061814  $10^{-3}$\\
  4 & 6.805008 $10^{-4}$ &7.465693  $10^{-4}$  & 8.116708 $10^{-4}$& 8.669160  $10^{-4}$& 9.131568 $10^{-4}$ \\
  3 & 5.356378 $10^{-4}$ & 5.968091 $10^{-4}$& 6.644138 $10^{-4}$& 7.185324 $10^{-4}$& 7.611445 $10^{-4}$ \\
  2 & 4.041484 $10^{-4}$ & 4.483612 $10^{-4}$& 5.143481 $10^{-4}$& 5.646143 $10^{-4}$& 6.032161 $10^{-4}$ \\
  1 & 2.577279 $10^{-4}$ & 2.849179 $10^{-4}$& 3.836672 $10^{-4}$& 4.161031 $10^{-4}$& 4.416124 $10^{-4}$ \\
  0 & & & 3.552849 $10^{-4}$& 3.963389 $10^{-4}$&4.154929 $10^{-4}$ \\
 -1 & & 2.624454 $10^{-4}$& 2.959146 $10^{-4}$& 3.393633 $10^{-4}$& 3.674454 $10^{-4}$ \\
 -2 & & 1.913175 $10^{-4}$& 2.557686 $10^{-4}$& 2.958986 $10^{-4}$& 3.2731776 $10^{-4}$ \\ 
 -3 & & 1.530947 $10^{-4}$& 2.218293 $10^{-4}$& 2.610756 $10^{-4}$& 2.9099566 $10^{-4}$ \\
 -4 & & 1.276013 $10^{-4}$& 1.938282 $10^{-4}$& 2.381342 $10^{-4}$& 2.6479426 $10^{-4}$ \\
 -5 & & 1.091393 $10^{-4}$& 1.703231 $10^{-4}$& 2.168287 $10^{-4}$& 2.5021866 $10^{-4}$ \\
 -6 & & 9.526346 $10^{-5}$& 1.511191 $10^{-4}$& 1.956943 $10^{-4}$& 2.3129086 $10^{-4}$ \\
 -7 & & 8.445787 $10^{-5}$& 1.354815 $10^{-4}$& 1.772801 $10^{-4}$& 2.1163136 $10^{-4}$ \\
 -8 & & 7.577619 $10^{-5}$& 1.225229 $10^{-4}$& 1.615688 $10^{-4}$& 1.9434676 $10^{-4}$ \\
 -9 & & 6.862877 $10^{-5}$& 1.116922 $10^{-4}$& 1.482299 $10^{-4}$& 1.7943816 $10^{-4}$ \\
-10 & & 6.266023 $10^{-5}$& 1.025184 $10^{-4}$& 1.367697 $10^{-4}$& 1.6645396 $10^{-4}$ \\
-11 & & 5.758742 $10^{-5}$& 9.466158 $10^{-5}$& 1.268847 $10^{-4}$& 1.5515566 $10^{-4}$ \\
-12 & & 5.325474 $10^{-5}$& 8.789878 $10^{-5}$& 1.182941 $10^{-4}$& 1.4523276 $10^{-4}$ \\
-13 & & 4.950474 $10^{-5}$& 8.199087 $10^{-5}$& 1.107265 $10^{-4}$& 1.3642856 $10^{-4}$ \\
-14 & & 4.622192 $10^{-5}$& 7.680159 $10^{-5}$& 1.040415 $10^{-4}$& 1.2859706 $10^{-4}$ \\
-15 & & 4.333618 $10^{-5}$& 7.221651 $10^{-5}$& 9.809704 $10^{-5}$& 1.2158186 $10^{-4}$ \\
-16 & & 4.077842 $10^{-5}$& 6.813154 $10^{-5}$& 9.278051 $10^{-5}$& 1.1527536 $10^{-4}$ \\
-17 & & 3.849929 $10^{-5}$& 6.447465 $10^{-5}$& 8.799700 $10^{-5}$& 1.0957106 $10^{-4}$ \\
-18 & & 3.645414 $10^{-5}$& 6.117758 $10^{-5}$& 8.366402 $10^{-5}$& 1.0438296 $10^{-4}$ \\
-19 & & 3.460764 $10^{-5}$& 5.819284 $10^{-5}$& 7.973133 $10^{-5}$& 9.9652696 $10^{-5}$ \\
-20 & & 3.293669 $10^{-5}$& 5.547975 $10^{-5}$& 7.614814 $10^{-5}$& 9.5325226 $10^{-5}$ \\
\hline 
\end{tabular}
\end{center}
\end{table*}
  
 Indeed the model without diffusion shows the largest errors. When the
sound speed is compared with  Basu's seismic model (see Fig. 1) it is seen 
that the discrepancy grows steadily from the center up to slightly below the
 convective envelope where it reaches $7.5 10^{-3}$. For $x=r/R <0.26$ the 
 sound speed is too large while it is too small in the rest of the model.
  Most of this error is suppressed when hydrogen diffusion is introduced.
Comparing Fig. 1 and 3, we see that part of the errors comes also from
differences in $\Gamma_1$  since otherwise errors in $u$ should just be twice
 these in $c$. In the radiative core our 
$\Gamma_1=\left(\frac{\partial\ln p}{\partial\ln \rho}\right)_S$ (where the 
subscript $S$ means that the derivative is taken at constant entropy)
are generally too small, 
 the difference is of $1.510^{-3}$ at the center and decreases steadily to 
 cancel at $x=0.67$. Then up to $x=0.9$, the absolute value of the differences 
 is a few times $10^{-4}$. These differences are of the same order as those
found with other equations of state as shown by Basu \& Christensen Dalsgaard
(1997). In the interior, these differences are mainly produced by errors in the 
evaluation of the Coulomb interactions. It may even be surprising that the simple 
Debye theory gives such accurate results. Its relative importance reaches a 
maximum at about $610^4$ K (Shibahashi et al. 1983) but as the temperature 
decreases another source of errors becomes more and more important and finally 
dominates. It comes from uncertainties in the ionization due in part to the 
deficiencies of the Debye theory but mainly to inaccuracies in the computation of 
the internal partition functions. All the errors accuring in the convective 
envelope have mainly a local influence and only a minor one on the global 
parameters since the entropy is constant there. The errors still present in the 
equation of state have little influence on the helium abundance and the other 
global parameters of the models as can be seen from the results of Morel et al. 
(1997) who have tested two EOS. As a matter of fact, the uncertainties in the 
diffusion theory are, in that respect, more important than those in the equation 
of state as changing its coefficients within the error bars gives larger 
variations of some of the global parameters such as Y 
(see Gabriel \& Carlier 1997).
  The density differences are also much larger than for the other models. This 
model also has a convective envelope which is too shallow and a surface $X$ 
value which is too low. Fig. 4 gives the comparison of the normalized frequencies 
(for the definition see Christensen-Dalsgaard \& Berthomieu 1991) of that
model with observations. It shows the 2 strips indicative of errors below the
convective envelope (see for instance Christensen-Dalsgaard \& Berthomieu 1991). 
 The upper one contains large degree modes confined to the
convective envelope while the lower one contains lower degree modes which have
also large amplitudes in the radiative core; modes of degree 30, 40 and 50 are 
seen crossing from the upper to the lower one  as they penetrate below the
envelope.  

 The model with hydrogen diffusion only is, among the first 4, that 
which compares best with the seismic model as the maximum discrepancy peaks  
to $1.6 10^{-3}$ a little below the convective envelope but is generally 
smaller than $10^{-3}$. The density profile is also among the best of this 
serie of models. The improvement comes partly because the error in $c$ 
becomes negative for $0.4 < x < 0,6$ but  the model  
shows a significant variation of the discrepancy around $x=0.6$ where it 
increases by $2.2 10^{-3}$. This might suggest that the gravitational 
settling is too strong there. The sound speed and $u$ in the central 
regions are now too small, the convective envelope is too deep but the 
surface $Y$ value (0.24434) is not far from Basu \& Antia values. 
These first 2 models are very close to those of Basu et al. (1996) with  
however differences such as the minimum in the sound speed differences, close
 to $x=0.2$. This minimum does not exist in our model 1 and is smaller in the 
second one. In this respect our model 1 is closer to model S1 of Morel et al. 
  (1997). The comparison of the theoretical frequencies with 
observations (see Fig. 5) shows a relatively good agreement though 2 close 
strips can be distinguished at high frequencies.
The results are also sensitive to the expression used for the diffusion 
coefficients, for instance if the formulae for $Ap$, $At$ and $Ax$ given by 
Thoul at al. (1994) are used, the fit is a little better. This implies that the 
remaining discrepancies are significantly influenced by the uncertainties in
the diffusion theory and Gabriel \& Carlier (1997) have checked that
modifications of the diffusion coefficients within their uncertainty can reduce
the discrepancies. 

Models 3 ($Z=0.02$) and 4 ($Z=0.021$) include gravitational settling  of the heavy 
elements. We find
that the surface $Z_S$ value has decreased by 10\% which is in good agreement
 with Proffitt \& Michaud (1991) but 2\% higher than found by Proffitt (1994) 
 and Morel et al. (1997). On the other hand our models have smaller central 
 $Z$ enrichment (close to 0.00085) than those of  Proffitt \& Michaud (0.0014),
  Proffitt (0.0012)  and Morel et al. (0.0009-0.001).  
  This can be explained by the different diffusion 
theories used and is indicative of their uncertainties.
The sound speed, $u$ and the density show larger discrepancies in the central 
regions and 
they increase with $Z$. For $0.2 < x < 0.6$ the discrepancies remain small but 
the slope is negative for $Z_0=0.021$ which means that $Z$ is a little too 
high there. Around $x \simeq 0.68$, the sharp increase of the discrepancy,
already noticed in model 2, is more pronounced and gives now a stronger bump. This 
bump  is found by everyone (see Morel et al. 1997; Basu 1997). 
 It shows that the depletion of $Z$ and the $X$ enrichment are too large.  
 In the convective envelope the models hardly change. 
 The convective zone of model 4 is too deep while that of
model 3 fits well the seismic values. The surface $Y$ values, respectively
0.24511 and 0.24933 are in good agreement with Basu \& Antia (1995) values.
Model 3 has a $Z/X$ value of 0.02444 very close to the spectroscopic value
while that of model 4 of 0.02587 is already a little larger. Fig. 6 and 7
show the comparison of the frequencies with observations. Model 3, with
$Z_0=0.02$, shows again the 2 strips though the gap is much narrower than for 
model 1. Model 4, with $Z_0=0.021$, gives a better fit as only one strip can 
be detected. However some modes of degree 30 and 40 can be seen below it. To
summarize, model 4 gives a better fit with the frequencies while model 3 shows
a better position of the envelope boundary and a $Z/X$ closer to the
spectroscopic value. 

One way to suppress the discrepancies below the convective envelope, is to 
increase the mixing in that region. 
Model 5 is obtained using Proffitt \& Michaud's formula. The bump close to 
$x=0.68$ is now replaced by an oscillation indicating too little rotationally 
induced mixing just below the envelope and too much deeper down.
To improve the situation, a slight modification of their formula has been
made. The formula (with Proffitt \& Michaud notations)
\[ \log D_T = \log F_s + \log D_0 +(q-q_*) 10^{1.16 + 0.48 \log a} \]
is used everywhere for $q > 0.5$ and $q_* =(q_e-0.01)$ (instead of 0.95) 
where $q_e$ is the mass fraction at the bottom of the convective envelope.
This gives model 6 which shows a better fit in that region. Its $c$ and $u$ 
show a positive slope in $ 0.2 < x< 0.6$ showing that $Z_0$ is a little too 
small.
Also the convective envelope base is close to the upper acceptable limit.
Figure 8 shows that the frequencies of model 6  are significantly better than
these of model 4.

To improve the fit further, we have computed  model 7 with $Z_0=0.0205$. To
have a good fit below the envelope we had to reduce $D_0$ by a factor 2.5 and 
$F_s$ is defined as
\[ 5min\{0.2,max[(\nabla_{ad}-\nabla),0.15]\}   \]
This model still shows
discrepancies in the sound speed relative to the seismic one larger than 
$1. 10^{-3}$ for $x < 0.1$, but in most of the radiative core they are of the
order of $6. 10^{-4}$; in the convective envelope they are close to 
$1. 10^{-3}$
as for all the models. Its convective envelope base is in good agreement with
the value of Basu \& Antia. However this model, as well as all those including
rotationally induced mixing, has indeed less surface hydrogen enrichment and
less heavy element depletion. As a result the surface $Y$ value of 0.2543 is
0.0025 higher than the Basu and Antia upper limit. Also, the $Z/X$ value 
(0.026258) is
close to the spectroscopic upper limit. The comparison of the frequencies of model
 7, given in Fig. 9, shows that they all fall in a narrow strip of less than 0.8
$\mu$Hz width indicating that most of the errors in the interior are removed. 
The slope at high frequencies is the signature of problem close to the surface.
Monteiro et al. (1994) have shown that the errors are decreased when using 
Canuto \& Mazitelli (1991) theory of convection while Gabriel (1995) has shown that 
the same result is obtained in the frame of the mixing length theory if this 
parameter increases with depth. This latter point of view has been confirmed 
by people who use Kurucz code (see for instance van't Veer-Menneret \& 
Megessier 1996; Schlattl et al. 1997). They find that a value of $l/H_p=0.5$ 
has to be used to reproduce the spectrum while a much large value is obtained 
through the computation of solar models. Also Demarque et al. (1997), find the 
same results using a variation of the mixing length suggested by numerical 
simulations of convection.

All models with gravitational settling show too large a central condensation.
This leads to small errors in the frequencies but the consequences can
 be seen better in the fine-structure spacing $d_l(n)=\nu_{l,n}-\nu_{l+2,n-1}$ 
 which is more sensitive to the structure of the central core.  Figures 10 and 11
give the differences between the theoretical and the observational values of 
$d_l(n)$ for $l=0$ and $1$ respectively. Taking into account the 
observational errors which are of the order of $0.2$ $\mu$Hz but varies from 
mode to mode (see Chaplin et al. 1997), we see that the model without diffusion 
shows larger errors but that it is presently impossible to distinguish between the others.
 
Finally, we give, in table 4, the low order p-mode and g-mode frequencies of 
model 7 for the degrees 0 to 4 (notice that the radial mode orders are increased 
by one, as usually done by observers). As some people are searching for high order 
g-modes in the data of some SOHO experiments, a list extending down to 
$\nu = 10^{-5}$ is available on request to gabriel@astro.ulg.ac.be.

\section{Conclusions}
The set of models presented here shows that diffusion improves definitively the
solar models. Surprisingly, when Fig. 1 to 3 are considered, the model with 
hydrogen diffusion only appears better than those taking also the heavy element 
diffusion into account but unfortunately, his convective zone is too deep. 
However the uncertainties in 
the diffusion theory influence significantly the remaining discrepancies with 
the seismic model. Nevertheless a problem is still affecting the layers below 
the convective envelope. We have shown that it could be resolved taking
rotationally induced mixing into account. However, even if this mechanism must 
be at work, others can also contribute to solve this difficulty. Gravity waves
 can have the same effect (Montalban 1994; Montalban \& Schatzman 1996; Kumar
\& Quataert 1997). As pointed out in the second section, the heavy elements are 
far from being fully ionized in these layers 
and their diffusion coefficients could significantly differ from the adopted 
values. But also, since our $p/\rho$ are too small, a magnetic contribution to 
the pressure could also help solving the problem.

Another problem can also be seen in the central core. 
This might suggest too much gravitational settling there. 
The flat rotation curve in the interior strongly suggest that rotationally 
induce mixing should be at work everywhere in the Sun. 
However there are other possibilities. Small opacity changes of only a few 
percents, can significantly change the models (Gabriel 1995, 1996;
Gabriel \& Carlier 1997). Also, maybe one should not forget that there is still 
another possibility of mixing as the Sun was found unstable when it was younger. 
There is some uncertainty on the duration of this unstable phase 
(for instance Boury et al. (1975) found that it lasts from $2.410^8$ to 
$3.10^9$ years) and its consequences are completely unknown but it could have 
produced some mixing and as a result have influenced  the later evolution of the 
Sun.

\end{document}